\documentclass[aps,amsmath,amssymb,prl,twocolumn,preprintnumbers,superscriptaddress,nobalancelastpage]{revtex4}
\usepackage{graphicx}
\newcommand{\be}{\begin{eqnarray}}
\newcommand{\ee}{\end{eqnarray}}

\begin{document}
\title{
Universal Scaling Relations in Strongly Anisotropic Materials
      }
\author{
M.~B.~Hastings
       }
\affiliation{
Center for Nonlinear Studies and Theoretical Division, Los Alamos National
Laboratory, Los Alamos, NM 87545, hastings@cnls.lanl.gov
            }
\author{
C.~Mudry
       }
\affiliation{
Paul Scherrer Institut, CH-5232 Villigen PSI, Switzerland
            }
\date{\today}
\begin{abstract}
We consider the critical temperature in strongly anisotropic antiferromagnetic
materials, with weak coupling between stacked planes, in order
to determine the interplane coupling constant from experimentally
measured susceptibilities.  We present theoretical
arguments for a universal relation between interplane coupling and
susceptibility shown numerically by 
Yasuda et.~al., Phys.~Rev.~Lett.~\textbf{94}, 217201 (2005).
We predict a more general scaling function if the system is close to a quantum
critical point, a similar relation for other susceptibilities than
considered in Yasuda et.~al., and the validity of these relations
for more general phase transitions.
\end{abstract}
\maketitle

Many materials display at low temperatures strongly 
spatially anisotropic responses to magnetic or electronic probes.
This fact has motivated the theoretical study of low dimensional 
quantum systems on their own right. Solving one- or two-dimensional
quantum systems
can be useful to understand intermediary regimes of temperature
in which fluctuations are dominated by the subsystem of lower
dimensionality. Three-dimensionality is effectively restored once 
the temperature is lowered below the lowest energy scale characterizing
the anisotropy. 

For magnetic systems this scale can be the temperature
$1/\beta^{\ }_{AF}$ ($k^{\ }_{B}=\hbar=1$)
below which antiferromagnetic (AF) long-range order manifests itself. 
In this context, one of the most studied model is perhaps
a stacking in three dimensions of chains or square lattices
on each of which a nearest-neighbor quantum spin-$S$ Heisenberg model
$H^{\ }_{J}$
with AF exchange coupling $J>0$ is defined.
To model a strong spatial anisotropy, 
one assumes that there exists a nearest-neighbor
AF exchange coupling $J'$ in the directions transverse
to the chains or planes that is much weaker than $J$,
$J\gg J'>0$.
The three-dimensional quantum Hamiltonian is $H^{\ }_{3d}$.
Many efforts have been invested for the last 30 years in
calculating the $J'$--dependence of 
$1/\beta^{\ }_{AF}$\cite{Scalapino75,Schulz96,Irkhin97,Bocquet02,Yasuda05}.
In this letter we shall address the question:
are there some universal relations that relate $J'$ and some observables
of $H^{\ }_{J}$ or $H^{\ }_{3d}$?

The motivation for this question comes from the work by
Yasuda et al.\ in Ref.\ \onlinecite{Yasuda05}
in which the N\'eel temperature $1/\beta^{\ }_{AF}$ of $H^{\ }_{3d}$,
the $n$-dimensional static staggered susceptibility
$
\chi^{(n)}_{s}=
\chi^{zz}(\boldsymbol{Q},\omega=0;\beta^{\ }_{AF})
$
of $H^{\ }_{J}$
where
$
\boldsymbol{Q}=
\pi
$
if $n=1$ or
$
\boldsymbol{Q}=(\pi,\pi)
$
if $n=2$,
and
$
J'\chi^{(n)}_{s}=
1/\zeta^{\ }_{n}(J')
$
were computed numerically as a function of $0<J'/J\leq 1$.
In quasi-two-dimension, it was observed that
\begin{eqnarray}
\label{eq: const}
J'\chi^{(n=2)}_{s}=1/\zeta^{\ }_{n=2}
\end{eqnarray}
becomes independent of $J'/J$ when $J'/J<0.1$
and that the constant value that it takes, although not 1/2
as predicted by mean-field theory,\cite{Scalapino75,Schulz96}
is independent of the magnitude of the spin and even takes the same
value for a classical ($S=\infty$) model.
Although the evidence is less pronounced
the same conclusion was reached in quasi-one-dimension.

We want to construct a tractable model that reproduces qualitatively 
these findings and we want to understand how these results can be useful 
to establish experimentally the implied universality.
We present a theoretical argument that as $J'/J\downarrow 0$, the function
$\zeta_{n=2}$ converges to a constant.  We make the following additional
predictions.  

First, we consider more general AF
models in the plane, and we consider the case in which, by tuning parameters,
it is possible to tune the planar model close to a quantum phase transition,
so that the zero-temperature AF order of the two-dimensional
model (without interplane couplings) becomes small.  Then, we predict the
scaling function
\begin{eqnarray}
\label{scale1}
J'\chi^{(2)}_{s}=
F^{\ }_1(c\beta^{\ }_{AF}/\xi^{(2)}),
\end{eqnarray}
for some scaling function $F^{\ }_1$,
in the limit $J'/J\downarrow 0$, where
$\xi^{(2)}$ is the correlation length of the two-dimensional model at
temperature $1/\beta^{\ }_{AF}$ and $c$ is a spin-wave velocity
defined below.  Note that in
the system considered by Yasuda et.~al.~the planar model is in
the renormalized classical regime so that 
$c\beta^{\ }_{AF}/\xi^{(2)}$ is exponentially
small in $c\beta^{\ }_{AF}$ and converges to zero as $J'/J\downarrow 0$.  
Therefore,
$F^{\ }_1(c\beta^{\ }_{AF}/\xi^{(2)}=0)=1/\zeta^{\ }_2$.

Second, we predict a similar scaling relation that will be valid for
quantities which are easier to access experimentally.  The susceptibility
$\chi^{(2)}_{s}$ defined above is that of the two-dimensional model without
the interlayer couplings, and cannot be measured in most real materials.
We define 
$\chi_{\pi,\pi,0}^{(3)}=\chi^{zz}(\pi,\pi,0,\omega=0;\beta^{\ }_{AF})$ 
to be the static susceptibility in the layered system
at wave vector $(\pi,\pi)$ in the plane and wave vector
$0$ perpendicular to the plane at temperature $1/\beta^{\ }_{AF}$.
Then, we predict that
\be
\label{scale2}
J'\chi_{\pi,\pi,0}^{(3)}=
F^{\ }_{2}(c^{(3)}\beta^{\ }_{AF}/\xi_{\pi,\pi,0}^{(3)}),
\ee
for some scaling function $F^{\ }_{2}$, in the limit $J'/J\downarrow 0$, 
where $c^{(3)}$ and $\xi_{\pi,\pi,0}^{(3)}$ are the in-plane spin-wave velocity
and correlation length of $H^{\ }_{3d}$ 
at temperature $1/\beta^{\ }_{AF}$ near wave vector $(\pi,\pi,0)$,
respectively. As it is the instantaneous structure factor $S^{\ }_{\pi,\pi,0}$
that is most readily measured~\cite{Ronnow99},
we note that the product $J' S^{\ }_{\pi,\pi,0} \beta_{AF}$
also should obey a scaling law of the form (\ref{scale2}) 
for some scaling function $F^{\ }_{3}$. 

Third, we predict that similar scaling results hold for other layered models.


In quasi one-dimension, we expect that similar scaling results will also hold.
This does not, however, help us understand the results of Yasuda et.~al.~in
quasi one-dimension.  
The scaling functions $F^{\ }_{1},F^{\ }_{2}$ imply that the classical
and quantum models will show the same $\zeta^{(1)}$ only if
$c\beta^{\ }_{AF}\ll\xi^{(1)}$.
However, as the one-dimensional Heisenberg model on a chain is not in the
renormalized classical regime but rather quantum critical, it should
have some non-zero $c\beta^{\ }_{AF}/\xi^{(1)}$ 
and should show a different $\zeta^{(1)}$ than the classical model.  
Thus, the one-dimensional results remain a puzzle.

{\it Physical Motivation---}
Here, we present a physical motivation for the results above and a
brief microscopic derivation of the relevant non-linear sigma model. 
In the next section, we show these scaling results using 
a renormalization group (RG) for this non-linear sigma model.
The reason for which we use this model
is that we want to illustrate the effects of field renormalization and
the non-linear sigma model RG
already has a nontrivial field renormalization at
leading order in the coupling constant,
while such a renormalization is not seen until
order $\epsilon^2$ ($1/N$) in a $4-\epsilon$ (large $N$) expansion.

Since the interplane interaction is weak, we can treat it perturbatively
at the microscopic level.  Following standard steps, in the
absence of the interplane interaction, we can first derive the partition
function for the two-dimensional $O(N)$ quantum non-linear sigma model 
($2d$QNLSM)
with field $\boldsymbol{n}^{\ }_{k}(\boldsymbol{r},\tau)$, 
where $k$ is a discrete index labelling individual planes,
$\boldsymbol{r}$ is a two-dimensional vector describing coordinates
in the plane, and $\tau$ is imaginary time. The relevant action for 
plane $k$ is
$S^{\ }_{k}=S^{(1)}_{k}+S^{(2)}_{k}$
where
\begin{subequations}
\label{eq: non-interacting planar QNLSM}
\begin{eqnarray}
S^{(1)}_{k}:=
\int\mathcal{L}^{(1)}_{k}\equiv
\int\limits_{0}^{\beta} d\tau
\int\limits_{a}^{L}d^{2}\boldsymbol{r}\,
\frac{c}{2ag}\left(\partial^{\ }_{\mu}\boldsymbol{n}^{\ }_{k}\right)^{2}
\label{eq: non-interacting planar QNLSM a}
\end{eqnarray}
and
\begin{eqnarray}
S^{(2)}_{k}:=
\int\mathcal{L}^{(2)}_{k}\equiv
-
\int\limits_{0}^{\beta} d\tau
\int\limits_{a}^{L}d^{2}\boldsymbol{r}\,
\frac{c}{a^{3}}
Z^{\ }_{h}
\,
\boldsymbol{h}
\cdot
\boldsymbol{n}^{\ }_{k}.
\label{eq: non-interacting planar QNLSM b}
\end{eqnarray}
\end{subequations}
Here, the lattice spacing $a$ plays the role of the
microscopic ultra-violet (UV) cutoff, 
i.e., $\Lambda\sim 1/a$ that of an upper cutoff on momenta.
The linear size $L$ of the plane is the largest length scale of
the problem.
The derivative 
$\partial_{\mu}=(\partial_{c\tau},\boldsymbol{\nabla})$
depends on the spin-wave velocity $c$ in the plane and is of order $Ja$. 
The dimensionless coupling constant $g$
depends on the microscopic details of the intraplane interactions.
The dimensionless background field 
$\boldsymbol{h}$, where $h=|\boldsymbol{h}|$, 
is the external source for a static staggered magnetic field conjugate 
to the planar AF order parameter 
of the underlying lattice model.
It breaks the $O(N)$
symmetry of Lagrangian (\ref{eq: non-interacting planar QNLSM a})
down to $O(N-1)$ and as such acts as an infra-red (IR) regulator.
The dimensionless coupling 
$Z^{\ }_{h}$
is the field renormalization constant associated to 
$\boldsymbol{n}^{\ }_{k}$.
The use of the continuum limit within each of the planes labelled by $k$ 
is justified if we are after the physics on length scales much longer than $a$.

The interplane nearest-neighbor AF coupling $J'$ gives
the characteristic interplane spin-wave velocity $c'\sim J'a$
and length scale $a'\equiv(J/J')^{1/2} a$.  
The couplings $J',g$ get renormalized as discussed below, 
so the velocity $c'$ changes at longer length scales.
For a very weak nearest-neighbor interplain AF coupling,
$J'\ll J$, the physics on length scales much larger than $a$ but yet
not much larger than $a'$ is captured by the partition function
\begin{subequations}
\label{eq: def SA3dQNLSM}
\begin{eqnarray}
&&
Z\!=\!
\int\limits_{\mathbb{R}^{N}}\!
\left[
\prod_{k}
\mathcal{D}[\boldsymbol{n}^{\ }_{k}]\,
\delta(\boldsymbol{n}^{2}_{k}-1)
\right]\!
\exp\!
\left(
-\!
\sum_{k}\!
\int\!
\mathcal{L}^{\ }_{k}\!
\right)\!,
\label{eq: def O(N) QNLSM for RG analysis a}
\\
&&
\mathcal{L}^{\ }_{k}=
\mathcal{L}^{(1)}_{k}
+
\mathcal{L}^{(2)}_{k}
+
\mathcal{L}^{(3)}_{k}.
\label{eq: def O(N) QNLSM for RG analysis b}
\end{eqnarray}
The Lagrangian $\mathcal{L}^{(3)}_{k}$ 
encodes the effect of the microscopic
nearest-neighbor interplane AF interaction $J'$.  To
compute this,
we use a Hubbard-Stratonovich transformation to replace the microscopic
interaction between any two spins in the cubic lattice with coordinates
$(i,j,k)$ and $(i,j,k+1)$ by an interaction of each spin
with a fluctuating magnetic field 
$\boldsymbol{H}^{\ }_{i,j,k}$.
When $J'\ll J$ and on intermediary length scales
$a\ll\lambda\lesssim a'$, the dominant mode for the
Hubbard-Stratonovich field is near momentum $(\pi,\pi)$ in the plane.
The action for the spins in the presence of this field is
the same as Eq.~(\ref{eq: non-interacting planar QNLSM b}) with
field $\boldsymbol{h}$ replaced by $\boldsymbol{H}^{\ }_{k}$.
In this approximation, integrating the
Hubbard-Stratonovich field 
gives the \textit{short-range} interplane interaction term
\be
\mathcal{L}^{(3)}_{k}=
\frac{J'Z'}{2a^{2}}
\left(
\boldsymbol{n}^{\ }_{k  }
-
\boldsymbol{n}^{\ }_{k+1}
\right)^2,
\ee
\end{subequations}
where $Z'$ renormalizes as $Z^{2}_{h}$ to lowest order in $J'/J$,
\be
Z'=
Z^{2}_{h}
\left[
1
+
\mathcal{O}(J'/J)
\right].
\label{eq: ward}
\ee
This defines the so-called three-dimensional strongly anisotropic $O(N)$
QNLSM ($3d$SAQNLSM).

This derivation of the $3d$SAQNLSM
considered interactions between spins positioned
directly above and below each other in neighboring planes.
It is possible to treat more complicated interplane interactions.
For example, going back to the cubic lattice, 
let there be an AF
interaction $J^{\ }_1$ between the spin at site $(i,j,k)$ with that at
$(i,j,k\pm 1)$ and
another AF interactions $J^{\ }_2$ to the spins at sites
$(i\pm 1,j,k\pm 1)$ and $(i,j\pm 1,k\pm 1)$.  Then, if $J^{\ }_1\ll J$ and
$J^{\ }_2\ll J$, we can derive the $3d$SAQNLSM with the interplane coupling 
$J'=J^{\ }_1-4J^{\ }_2$.

\textit{Renormalization Group---}
Here, we present an RG analysis of the non-linear sigma model
(\ref{eq: def SA3dQNLSM}).
The most important result in this section is that the identity
(\ref{eq: ward})
is preserved under the RG flow up to the length scale
at which the scale dependent effective anisotropy 
(\ref{eq: flow of alpha if strong ani})
is of order 1.

\begin{figure}[!ht]
\centering
\includegraphics[width=1\columnwidth]{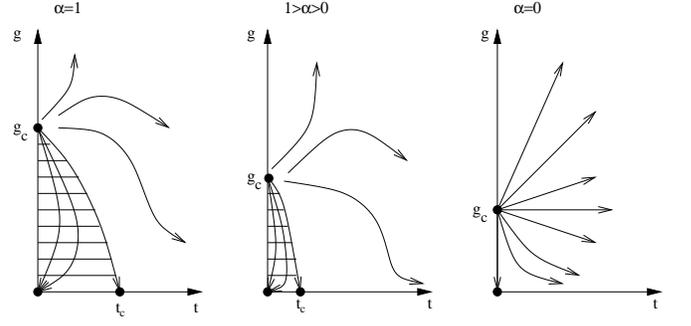}
\caption{
\label{fig: phase_dia_1}
Phase diagram and RG flows for the $3d$QNLSM $(\alpha=1$) 
and $2d$QNLAM $(\alpha=0$) after Ref.~\cite{CHN89}. 
Conjectured phase diagram and RG flow for the anisotropic $3d$QNLSM
($1>\alpha>0$). The shaded regions have long range N\'eel order.
The figure for $1>\alpha>0$ shows a two-dimensional slice of the
three-dimensional RG flow, as $\widetilde{\alpha}$ is changing under this
flow.
        }
\end{figure}

We perform a RG analysis following
Polyakov for convenience \cite{Polyakov75}.
In each plane labelled by $k$, we write 
\be
\boldsymbol{n}^{\ }_{k}&=&
\boldsymbol{m}^{\ }_{k}
\left(1-\boldsymbol{\phi}^{2}_{k}\right)^{1/2} 
+
\sum_{a=1}^{N-1} 
\boldsymbol{e}^{a}_{k}
\phi^{a}_{k}.
\label{eq: polyakov parametrization of spin waves}
\ee
The field of unit length $\boldsymbol{m}^{\ }_{k}$
encodes the planar AF order 
expected in the limit $g/c\downarrow0$,
while the $N-1$ fields $\boldsymbol{e}^{a}_{k}$
capture the deviations away from the direction 
$\boldsymbol{m}^{\ }_{k}$ 
of symmetry breaking,
i.e., the
$N-1$ $\boldsymbol{e}^{a}_{k}$
form an orthonormal basis of vectors orthogonal to $\boldsymbol{m}^{\ }_{k}$.
The $N-1$ coefficients $\phi^{a}_{k}$
make up the vector $\boldsymbol{\phi}^{\ }_{k}$.
To leading order in an expansion in powers of $g/c$ of
the parametrization (\ref{eq: polyakov parametrization of spin waves}),
the field $\boldsymbol{m}^{\ }_{k}$ is the slow mode while the $N-1$ fields
$\phi^{a}_{k}$ represent fast modes with 
characteristic 2-momenta 
$\widetilde{\Lambda}<|\boldsymbol{p}|\leq \Lambda$.
Substituting Eq.~(\ref{eq: polyakov parametrization of spin waves})
into
Eq.~(\ref{eq: def SA3dQNLSM}) 
gives the Lagrangian
\be
&&
\mathcal{L}^{(1)}_{k}
+
\mathcal{L}^{(2)}_{k}=
\frac{c}{2ag}
\left[
\left(
\partial^{\ }_{\mu}\phi^{a}_{k}
-
A^{ab}_{k\mu}\phi^{b}_{k}
\right)^2
+
\left(B^{a}_{k\mu}\right)^2
\right.
\\
\nonumber
&&
\left.
+
B^{a}_{k\mu}B^{b}_{k\mu}
\left(
\phi^{a}_{k}\phi^{b}_{k}
-
\boldsymbol{\phi}^{2}_{k}\delta^{ab}
\right)
\right]
-
\frac{c}{a^{3}}
Z^{\ }_h 
\boldsymbol{h}
\cdot
\boldsymbol{m}^{\ }_{k}
(1-\boldsymbol{\phi}^{2}_{k})^{1/2}
\label{eq: Polyakov parametrization of L(1)+L(2)}
\ee
to leading order in an expansion in powers of $g/c$.
The $N-1$ coefficients $B^{a}_{k\mu}$ 
are defined by
$
\partial^{\ }_{\mu}\boldsymbol{m}^{\ }_{k}=
\sum_{a=1}^{N-1} B^{a}_{\mu}\boldsymbol{e}^{a}_{k}
$.
The $(N-1)(N-2)/2$ independent coefficients 
$A^{ab}_{k\mu}=
-(\partial^{\ }_{\mu}\boldsymbol{e}^{b}_{k})\cdot\boldsymbol{e}^{a}_{k}
$. 
The RG flows of the dimensionless couplings
$g$,
$Z^{\ }_{h}h$,
and
$t\equiv 1/(J\beta)$
that follow after integration over the fast modes 
$\boldsymbol{\phi}^{\ }_{k}$
in the limit of no interplane interactions
were computed by Chakravarty, Halperin, and Nelson
to leading order in $g/c$ (see Fig.~\ref{fig: phase_dia_1}) \cite{CHN89}.
To this order, $c$ is unchanged.

To quantify the very weak microscopic interplanar coupling,
we define the anisotropy $\widetilde{\alpha}$
as the ratio of the importance of 
$\mathcal{L}^{(1)}_{k}$ 
to 
$\mathcal{L}^{(3)}_{k}$
when the upper cutoff on the momenta is $\widetilde{\Lambda}$.
By assumption, this anisotropy is strong
at the microscopic level (upper cutoff $\Lambda$),
\begin{eqnarray}
\alpha=
gJ'Z'/J\ll
1. 
\label{eq: def micro alpha}
\end{eqnarray}
Next, we consider the renormalization of 
$\mathcal{L}^{(1)}_{k}$ in 
Eq.~(\ref{eq: Polyakov parametrization of L(1)+L(2)})
when $\boldsymbol{h}=0$
and of the interplane interaction
\begin{eqnarray}
\mathcal{L}^{(3)}_{k}&=&
\frac{J'Z'}{2a^{2}}
\left[
\left(1-\boldsymbol{\phi}^{2 }_{k}\right)^{1/2}
\boldsymbol{m}^{\ }_{k}
-
\left(1-\boldsymbol{\phi}^{2 }_{k+1}\right)^{1/2}
\boldsymbol{m}^{\ }_{k+1}
\vphantom{\sum_{a=1}^{N-1}}
\right.
\nonumber\\
&&
\left.
+
\sum_{a=1}^{N-1}
\left(
\phi^{a}_{k}\boldsymbol{e}^{a}_{k}
-
\phi^{a}_{k+1}\boldsymbol{e}^{a}_{k+1}
\right)
\right]^{2}
\end{eqnarray}
after averaging over the fast modes $\boldsymbol{\phi}^{\ }_{k}$.
To this end, we shall introduce the renormalized values
\begin{subequations} 
\label{eq: RG of L1 if strong ani}
\begin{eqnarray}
\frac{1}{\widetilde{g}}&=&
\frac{1}{g}
\left(\frac{\Lambda}{\widetilde{\Lambda}}\right)
\left(
1
+
\left\langle
\phi^{a}_{k}\phi^{b}_{k}
-
\boldsymbol{\phi}^{2}_{k}\delta^{ab}
\right\rangle
\vphantom{\frac{1}{2}}
\right),
\label{eq: flow 1/g if strong ani}
\\
\frac{1}{\widetilde{t}}&=&
\frac{1}{t}
\left(
1
+
\left\langle
\phi^{a}_{k}\phi^{b}_{k}
-
\boldsymbol{\phi}^{2}_{k}\delta^{ab}
\right\rangle
\vphantom{\frac{1}{2}}
\right),
\label{eq: flow 1/t if strong ani}
\\
\widetilde{Z}^{\ }_h&=&
Z^{\ }_h 
\left(
1
-
\frac{1}{2}
\left\langle
\boldsymbol{\phi}^{2}_{k}
\right\rangle
+\cdots
\right),
\label{eq: flow Zofh if strong ani}
\end{eqnarray}
\end{subequations} 
at the scale $\widetilde{\Lambda}$
as a result of averaging over the fast modes 
$\boldsymbol{\phi}^{\ }_{k}$. 
For $\alpha\ll 1$, 
this average over fast modes is
$
\langle 
\phi^{a}_{k}
\phi^{b}_{l}
\rangle=
\delta^{\ }_{kl}
\delta^{ab}
\ln(\Lambda/\widetilde{\Lambda})
\frac{g}{4\pi}\coth(g/2t)
$.
Furthermore, 
\begin{eqnarray}
\mathcal{L}^{(3)}_{k}\approx
\frac{J'Z'}{2a^{2}}
\frac{\widetilde{Z}^{2 }_{h}}{Z^{2 }_{h}}
\left(
\boldsymbol{m}^{\ }_{k  }
-
\boldsymbol{m}^{\ }_{k+1}
\right)^2
\end{eqnarray}
from which follows the renormalizations
\begin{eqnarray}
\widetilde{\alpha}&=&
\left(\frac{\Lambda}{\widetilde{\Lambda}}\right)^{3}
\frac{\widetilde{g}\widetilde{Z'}}{gZ'}\,
\alpha,
\qquad
\frac{\widetilde{Z'}}{Z'}=
\frac{\widetilde{Z}^{2}_{h}}{Z^{2}_{h}},
\label{eq: flow of alpha if strong ani}
\end{eqnarray}
so long as $\widetilde\alpha\ll 1$.

Let us follow the RG flows encoded by
Eqs.~(\ref{eq: RG of L1 if strong ani})
and
Eqs.~(\ref{eq: flow of alpha if strong ani})
starting from the initial values
$g>0$, 
$t\equiv1/(J\beta^{\ }_{AF})<\infty$, 
and $1\gg\alpha>0$,
see Eq.~(\ref{eq: def micro alpha}),
corresponding to a point on the phase boundary between the 
N\'eel and paramagnetic phase (see Fig.~\ref{fig: phase_dia_1}).
Aside from the thermal de Broglie wavelength of the spin waves
$c\beta^{\ }_{AF}$,
the initial values $g$ and $t$ define a second characteristic
length scale, the correlation length $\xi^{(2)}$ in the $2d$QNLSM,
in view of $1\gg\alpha>0$. We shall distinguish two cases. 
In the renormalized classical regime $c\beta^{\ }_{AF}/\xi^{(2)}\ll1$.
In the quantum critical regime $c\beta^{\ }_{AF}/\xi^{(2)}\sim1$.
Finally, we denote by $\xi^{\ }_{\mathrm{cross}}$
the RG length scale at which $\widetilde{\alpha}\sim1$
and beyond which the RG flows
Eqs.~(\ref{eq: RG of L1 if strong ani})
and
Eqs.~(\ref{eq: flow of alpha if strong ani})
should be replaced by the flows of the isotropic $3d$QNLSM; naive
scaling gives $\xi_{\mathrm{cross}}\sim a'$, but the
RG flows above will change this scaling.
Any two of these characteristic length scales, $c\beta_{AF}, \xi^{(2)},$
and $\xi_{\mathrm{cross}}$, fix the third one
since the RG flows are constrained to the boundary between
the N\'eel and paramagnetic phases by assumption.
Without loss of generality, we shall consider the case
$\Lambda>1/(c\beta^{\ }_{AF})$.
As we lower the upper momentum cutoff,
the RG scale $\widetilde{\Lambda}^{-1}$ will  
eventually become larger than $c\beta^{\ }_{AF}$.
We shall consider RG scales 
$\widetilde{\Lambda}^{-1}\gg c\beta^{\ }_{AF}$, 
for which the quantum fluctuations are important.

We begin with the renormalized classical regime
of the $3d$SAQNLSM. As is illustrated in Fig.~\ref{fig: phase_dia_1},
the running coupling constants 
$\widetilde{g}$ in Eq.~(\ref{eq: flow 1/g if strong ani}) and
$\widetilde{t}$ in Eq.~(\ref{eq: flow 1/t if strong ani}) 
flow towards zero and $\infty$, respectively,
as $\widetilde{\Lambda}$ decreases but so long 
$\widetilde{\Lambda}^{-1}\lesssim c\beta^{\ }_{AF}$.
By Eq.~(\ref{eq: flow of alpha if strong ani}), 
the effective anisotropy decreases; 
using naive scaling which is
valid for $\tilde g\ll1$, we have 
$\widetilde{\alpha}\sim (\Lambda/\tilde \Lambda)^2 \alpha$.
Beyond  the RG length scale 
$\widetilde{\Lambda}^{-1}\sim c\beta^{\ }_{AF}$
we can replace the $2d$QNLSM in each plane by a classical $2d$NLSM 
with the effective coupling
$\widetilde{g}^{\ }_{cl}$, where $\widetilde{g}^{\ }_{cl}=\widetilde{g}$
at the scale $\widetilde{\Lambda}^{-1}\sim c\beta^{\ }_{AF}$.
The effective anisotropy of the classical
$3d$SANLSM continues to decrease as 
$
\widetilde{\alpha}=
(\Lambda/\widetilde{\Lambda})^2 
(\Lambda c\beta^{\ }_{AF}) 
[\widetilde{g}^{\ }_{cl}\widetilde{Z'}/(gZ')] 
\alpha=(\Lambda/\widetilde{\Lambda})^2 \widetilde{g}^{\ }_{cl} \widetilde{Z'}
J' \beta^{\ }_{AF}
$
continues to grow until it reaches the isotropic
RG scale $\widetilde{\alpha}\sim1$.
Equation~(\ref{eq: ward}) can then no longer hold.  
However, since there is only a finite range of scales over which 
$\widetilde \alpha$ is non-negligible but still less than unity,
we deduce that, at the scale $\widetilde{\alpha}\sim1$,
$\widetilde{Z'}\sim\widetilde{Z}^{2}_{h}$, 
up to some constant of order unity.  
Furthermore, $\widetilde{g}^{\ }_{cl}\sim 1$
at this scale also since it lies at some point on the phase boundary
between the N\' eel and paramagnetic phases. But
$\widetilde{g}^{\ }_{cl}\sim 1$
tells us that the corresponding RG scale $\widetilde{\Lambda}$
is of the order of the correlation length of the $2d$QNLSM.
In turn, this allows us to infer that the static staggered susceptibility of 
the $2d$QNLSM is given by
$
\chi^{(2)}_{s}\sim
(\Lambda/\widetilde{\Lambda})^2 \widetilde{Z}^{2}_{h}\beta^{\ }_{AF}
$,
up to universal corrections of order unity.  
We now multiply 
$\chi^{(2)}_{s}$
by
$J'$
estimated from the anisotropy $\alpha$ of the classical $3d$SANLSM,
$
J'\chi^{(2)}_{s}\sim 
\widetilde{\alpha}
(\widetilde{Z}^{2}_{h}/\widetilde{Z'})/\widetilde{g}^{\ }_{cl}
$.
Using Eq.~(\ref{eq: ward}), and the fact that
$\widetilde{g}^{\ }_{cl}\sim 1$ and $\widetilde\alpha\sim 1$, we
arrive at Eq.~(\ref{eq: const}) in the renormalized classical regime.

Note that each of these relations, such as $\widetilde\alpha\sim 1$ and
$\widetilde{g}^{\ }_{cl}\sim 1$, is defined up to a 
multiplicative constant that depends on the details of how 
we define the RG. 
However, the dimensionless combination in Eq.~(\ref{eq: const}) is
universal.  The reason for the universality is that all the microscopic
details of the Heisenberg model are encoded into the three independent
quantities
$\widetilde{g}$, 
$\widetilde{Z'}$, 
and 
$\widetilde{Z}^{\ }_h$
on any length scale much larger than $a$.
Let us perform the RG flow to some scale such that
$\widetilde{\alpha}$ is much less than unity.  Then, the identity
(\ref{eq: ward}) relates $\widetilde{Z'}$ to $\widetilde{Z}^{\ }_h$, 
leaving only two quantities independent in the classical regime, say
$\widetilde{g}^{\ }_{cl}$ and $\widetilde{\alpha}$.
The requirement of criticality relates $\widetilde{g}^{\ }_{cl}$
to $\widetilde{\alpha}$, leaving only one independent quantity, 
say $\widetilde{\alpha}$. 
Choosing the renormalization scale to be some given
fraction of the correlation length in the two-dimensional model 
fixes the last quantity, and thus there are no independent parameters left.

Near a quantum critical point and as is the case 
for the renormalized classical regime,
the length scale at which $\widetilde{\alpha}\sim1$
is of the order $\xi^{(2)}$. Now, however,
there is no significant separation of scales between 
$c\beta^{\ }_{AF}$ and $\xi^{(2)}$ anymore, i.e., 
$\widetilde{\alpha}\sim1$
already at $c\beta^{\ }_{AF}$.
Correspondingly,
there will be universal corrections to (\ref{eq: ward})
in the form 
$\widetilde{Z'}/Z'=\kappa(1,g(1))\widetilde{Z}^{2}_{h}/Z^{2}_{h}$
where the function $\kappa$ of $\alpha$ and $g$
is universal with $\kappa(0,g)=1$.
The deviations in the quantum critical regime
from the limiting value of $J'\chi^{(2)}_{s}$
in the classical renormalized regime define 
the universal scaling function
$F^{\ }_{1}$ of $c\beta^{\ }_{AF}/\xi^{(2)}$.

Similarly, the correlation
at the $(\pi,\pi,0)$ point is of order $\widetilde{\Lambda}^{-1}$ 
in the plane while it is of order a single
interlayer spacing between the planes.  
Thus, 
$
\chi_{\pi,\pi,0}^{(3)}\sim
\chi^{(2)}_{s}
$ 
and $\xi_{\pi,\pi,0}^{(3)}\sim \xi^{(2)}$, 
and so Eq.~(\ref{scale2}) follows.

We close by noting that all arguments presented here for 
a non-linear sigma model with $O(N)$ symmetry extend to
non-linear sigma models defined on Riemannian manifolds with
a positive curvature tensor.
For example, we expect similar universal scaling relations
for a stacking of AF Heisenberg models on a triangular lattice.

\textit{Discussion---}
We have provided a field-theoretic basis for understanding the result of
Yasuda et.~al in the quasi-two-dimensional case,
generalized it to deal with situations near quantum critical points, 
and given a version thereof expressed in terms of experimentally accessible 
quantities. The calculation of the scaling functions 
$F^{\ }_{1,2,3}$ within a field theoretic
approach will only be approximate, and the best estimate for $F^{\ }_{1}(0)$ 
is given by numerical calculations. 
For example, it can be shown that
the mean-field result $F^{\ }_{2}(0)=1/2$ follows from the large $N$ limit
of the $O(N)$ $3d$SAQNLSM. 
It would be very valuable to perform
a numerical study of $F^{\ }_{2,3}$, in both the renormalized classical
and quantum critical regimes.

{\it Acknowledgements---}
We thank M. Troyer for explaining \cite{Yasuda05}
and H.~M.~Ronnow for useful discussions.
This work was supported by DOE grant 
W-7405-ENG-36.  

\end{document}